\documentclass[aps,pra,twocolumn,showpacs,showkeys]{revtex4-1}

\usepackage{graphicx}
\usepackage{amsthm,amssymb,amsmath}
\usepackage{textcomp}
\usepackage{xspace}
\usepackage{color}

\begin{document}


\title{A universal, plug-and-play synchronisation scheme for practical quantum networks}

\author{Virginia D'Auria}
\email{Virginia.DAuria@univ-cotedazur.fr}
\affiliation{Universit\'e C\^ote d'Azur, CNRS, Institut de Physique de Nice (INPHYNI), UMR 7010, Parc Valrose, 06108 Nice Cedex 2, France.}
\author{Bruno Fedrici}
\affiliation{Universit\'e C\^ote d'Azur, CNRS, Institut de Physique de Nice (INPHYNI), UMR 7010, Parc Valrose, 06108 Nice Cedex 2, France.}
\author{Lutfi Arif Ngah}%
\affiliation{Universit\'e C\^ote d'Azur, CNRS, Institut de Physique de Nice (INPHYNI), UMR 7010, Parc Valrose, 06108 Nice Cedex 2, France.}
\author{Florian Kaiser}%
\email{Present  address: 3$^{rd}$ Institute of Physics, University of Stuttgart and Institute for Quantum Science and Technology IQST, Germanyfaffenwaldring 57, 70569 Stuttgart, Germany.}
\affiliation{Universit\'e C\^ote d'Azur, CNRS, Institut de Physique de Nice (INPHYNI), UMR 7010, Parc Valrose, 06108 Nice Cedex 2, France.}
\author{Laurent Labont\'e}%
\affiliation{Universit\'e C\^ote d'Azur, CNRS, Institut de Physique de Nice (INPHYNI), UMR 7010, Parc Valrose, 06108 Nice Cedex 2, France.}
\author{Olivier Alibart}%
\affiliation{Universit\'e C\^ote d'Azur, CNRS, Institut de Physique de Nice (INPHYNI), UMR 7010, Parc Valrose, 06108 Nice Cedex 2, France.}
\author{S\'ebastien Tanzilli}%
\affiliation{Universit\'e C\^ote d'Azur, CNRS, Institut de Physique de Nice (INPHYNI), UMR 7010, Parc Valrose, 06108 Nice Cedex 2, France.}

\date{\today}

\begin{abstract}
We propose and experimentally demonstrate a plug-and-play, practical, and enabling method allowing to synchronize the building blocks of a quantum network in an all-optical way. Our scheme relies on mature and reliable classical telecommunication and non-linear optical technologies and can be implemented in a universal way with off-the-shelf components. Compared to already reported solutions, it allows achieving high-quality synchronization compatible with high network-operation rate and is free from opto-electronic jitters affecting servo-loop based configurations. We test our scheme with a genuine quantum optical method in terms of the interference between two photons coming from two remotely synchronized sources spaced by distances of up to 100\,km. Measured visibilities well above 90\% confirm the validity of our approach. Due its simplicity and high-quality performance, our scheme paves the way for the synchronization of long-distance quantum networks based on fibre, free-space, as well as hybrid solutions.
\end{abstract}



\maketitle

\section*{Introduction}

\subsection*{Quantum networks and synchronization issues}

In the context of digital society, quantum networks promise to combine highly efficient data processing with ultra-secure data exchanges~\cite{Kimble2008,Simon2017,Wehner2018}. This vision has motivated the development of crucial constituents including quantum memories~\cite{Saglamyurek2015,Parigi2015}, coherent interfaces between different systems~\cite{Woerkom2018,Maring2017,Kaiser2015,Kaiser2019}, and optical quantum communication links based on quantum teleportation over long fibre connections~\cite{Sun2017,Sun2016,Valivarthi2016,Jin2015}. Nevertheless, the development of operational quantum networks remains hindered by the lack of practical synchronisation methods that allow the different building blocks to work together under high timing accuracy. If different synchronization strategies have been investigated both in pulsed and in continuous wave regimes, they suffer from important limitations in terms of achievable operation rates and distances~\cite{Xia2018,Tao2006,Halder2007,Alibart2016}. The pulsed regime offers the advantage of introducing an intrinsic time-binning. However, in most of reported realizations, synchronization has demanded the implementation of sophisticated, and hardly scalable, phase-locked loops~\cite{Tao2006,Kaltenbaek2009} or atomic clocks and dedicated electronics~\cite{Shelton2001,Cundiff2003}. As relevant and recent examples, two out-of-the-laboratory demonstrations~\cite{Sun2016,Valivarthi2016} have established the state-of-the-art for discrete variable quantum networks, by showing entanglement teleportation across metropolitan areas. In both schemes, the synchronization of the network nodes relies on custom feedback systems. These latter, as for all strategies relying on servo-loops~\cite{Tao2006,Halder2007,Alibart2016,Kaltenbaek2009}, are based on opto-electronic conversions that come at the price of unavoidable timing-jitters. To mitigate their effect, photons carrying the quantum information must be filtered so as to have long coherence times and relax the constraints on the synchronization. This requires the use of spectral filtering stages~\cite{Halder2007} and, more importantly, poses a fundamental limitation to the maximum clocking regime in order to avoid signals from different photons to superpose. Alternatively, a different approach is based on continuous operation regime and on the post-selection of single time mode events by means of detection~\cite{Halder2007,Halder2008}. In this case, as no intrinsic time-binning is provided by the continuous lasers, detector timing jitters become the principal source of time uncertainty. However, to comply with typical detection jitters (at best few tens of ps)~\cite{Hadfield2009}, extremely narrow spectral filtering stages are demanded with a dramatic effect on working rate.

In this study, we consider the pulsed regime and show that an efficient, universal, and plug-and-play solution to the quantum network synchronisation is provided by the combination of off-the-shelf classical telecommunication and nonlinear optics technologies. Our idea is to exploit classical telecom know-how to distribute an all-optical clock signal over a fibre network and to apply locally-adapted nonlinear optical stages~\cite{Alibart2016} to tailor its spectral properties and make it drive directly the devices located at the different nodes. By using suitable nonlinear optical conversions, this scheme can be employed to deliver a common time-reference to distant quantum systems with no a priori restrictions on either their nature or number. We successfully tested it by synchronizing two distant photon pair sources. Compared to other reported solutions, our fully-optical method offers high synchronisation quality without the need for complex clock or control systems. This strategy dramatically reduces the experimental overhead and bypasses accuracy limitations due to finite speed~\cite{Kaltenbaek2009} and timing-jitters of current servo-loop systems~\cite{Abooussouan2010}.

Its extreme simplicity, universality, and feasibility with current technologies make our synchronisation scheme a valuable strategy for future, long distance, quantum networks made with fibres and/or in-space architectures.

\subsection*{A timing-jitter free solution}

In a practical situation, the master clock is represented by a pulsed laser emitting at a telecom wavelength compatible with long distance distribution in optical fibres. Classical telecom technology represents a particularly convenient work frame by providing ultra-fast optical clocks with pulse repetition rates up to 10\,GHz~\cite{Ngah2015}, chromatic dispersion compensation modules, high gain erbium-doped fibre amplifiers (EDFA), and low loss fibre components, including dense wavelength demultiplexing stages (DWDM)~\cite{Aktas2016}.

\begin{figure}
\includegraphics[width=0.5\textwidth]{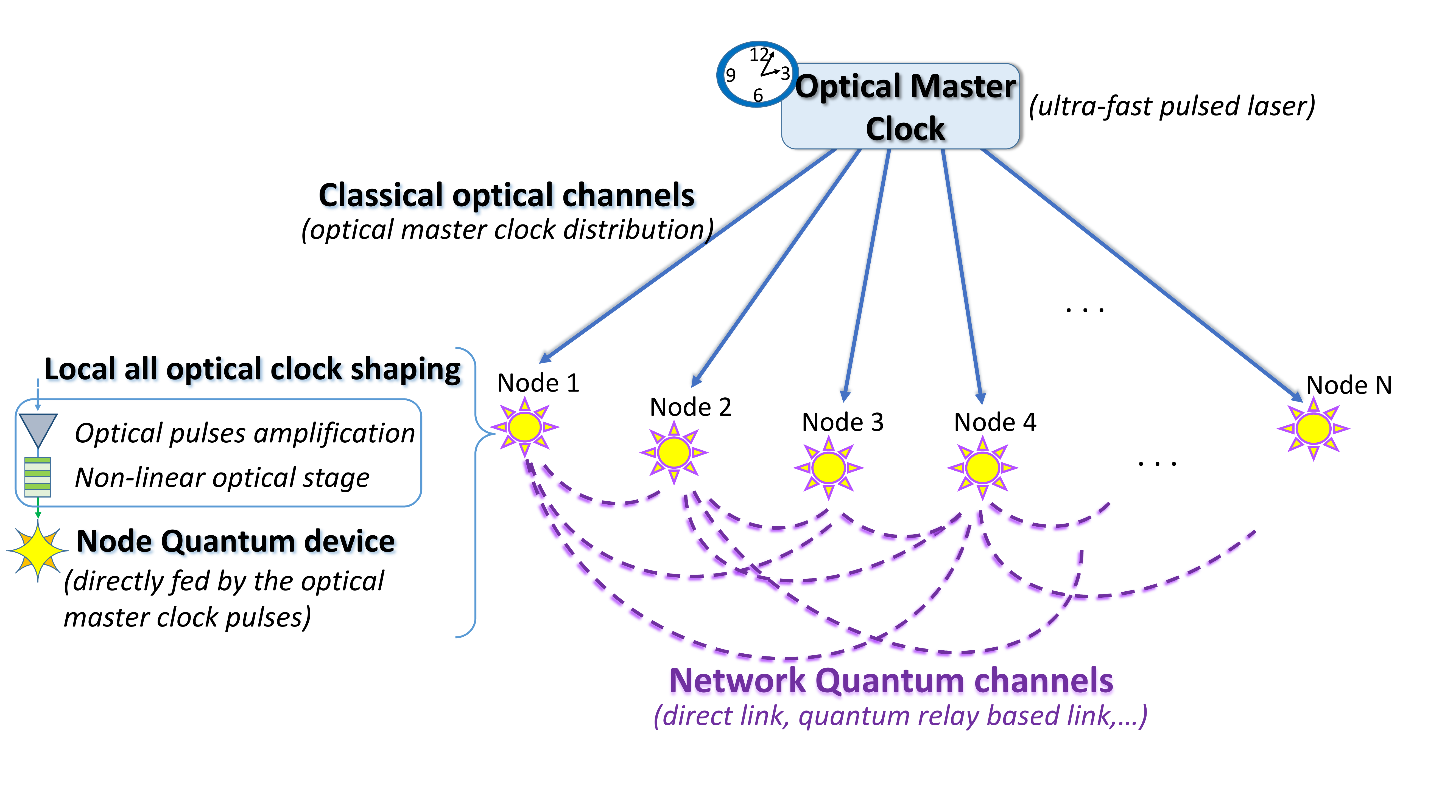}
\caption{Conceptual architecture of our synchronisation procedure based on an all-optical master clock. Optical clock laser pulses are distributed in parallel to different nodes of the quantum network via classical optical channels. At each of the nodes, depending on the specific situation, the clock pulses are locally shaped in an all-optical fashion via amplification and nonlinear optical stages and used to feed directly the quantum devices (entanglement generation or storage) present at the node. By doing so, the nodes driven by the same optical clock are automatically synchronized without the introduction of any jitters and without the need for control systems. As for general quantum networks, they can now be connected by quantum communication channels and used to share quantum information.}
\label{figprinciple}
\end{figure}

As shown in \figurename{~\ref{figprinciple}}, master clock pulses are distributed in parallel to different network nodes and are locally used to optically pump remote quantum devices. By doing so, the quantum systems (nodes 1 to n in \figurename{~\ref{figprinciple}}) are all driven by synchronous optical pulses and are, as a consequence, automatically provided with the same intrinsic timing reference. We stress that, since the clock laser does not convey any quantum information, its routing over the network can fully benefit from standard classical telecommunication tools. Undesired nonlinear effects caused by the propagation of ultra-short optical pulses over long optical fibres~\cite{Agrawal2006} can be overcome by using Fourier-transform-limited optical pulses lasting a few picoseconds (instead of femtoseconds). At the same time, advances in nonlinear optics ensure that off-the-shelf components can be employed, if necessary, to efficiently shape the master clock signal in the frequency domain so as to make it compatible with a specific and operational quantum device~\cite{Alibart2016}. This includes the possibility, after the distribution, of obtaining pulses that are longer than picosecond by means of adequate local spectral filtering.

Conceptually, the entire process of clock sharing, and manipulation is all-optical and free from parasite timing-jitters, typically in the order of tens of ps~\cite{Xia2018,Abooussouan2010}, that arise from opto-electronic conversions in servo-loops. In this regard, we emphasise that random timing-jitters are responsible for the main limitation in timing precision for most of the reported synchronisation configurations. Moreover, they represent an intrinsic limit to maximum achievable clock rates as they cause signals from different clock cycles to superpose~\cite{Hadfield2009}. In our scheme, the timing accuracy is in principle only defined by the stability of the master clock repetition rate, whose uncertainty is well below 100\,fs for off-the-shelf ultra-fast lasers and can go down to a fraction of a femtosecond in research-grade systems~\cite{Benedick2012}. Accordingly, our method guarantees high precision and is straightforwardly compatible with ultra-fast operation driven by master clock lasers paced at GHz repetition rates~\cite{Ngah2015}. Eventually, in the particular case of quantum network configurations requiring an interferometric phase stabilisation~\cite{Rozpedek2019}, our method is fully compliant with the addition of an optical frequency reference, as achieved in large infrastructures for time-frequency dissemination~\cite{Lisdat2016}.

\section*{Results \& Discussion}

\subsection*{Synchronisation tests on short distances}

\begin{figure*}
\includegraphics[width=0.96\textwidth]{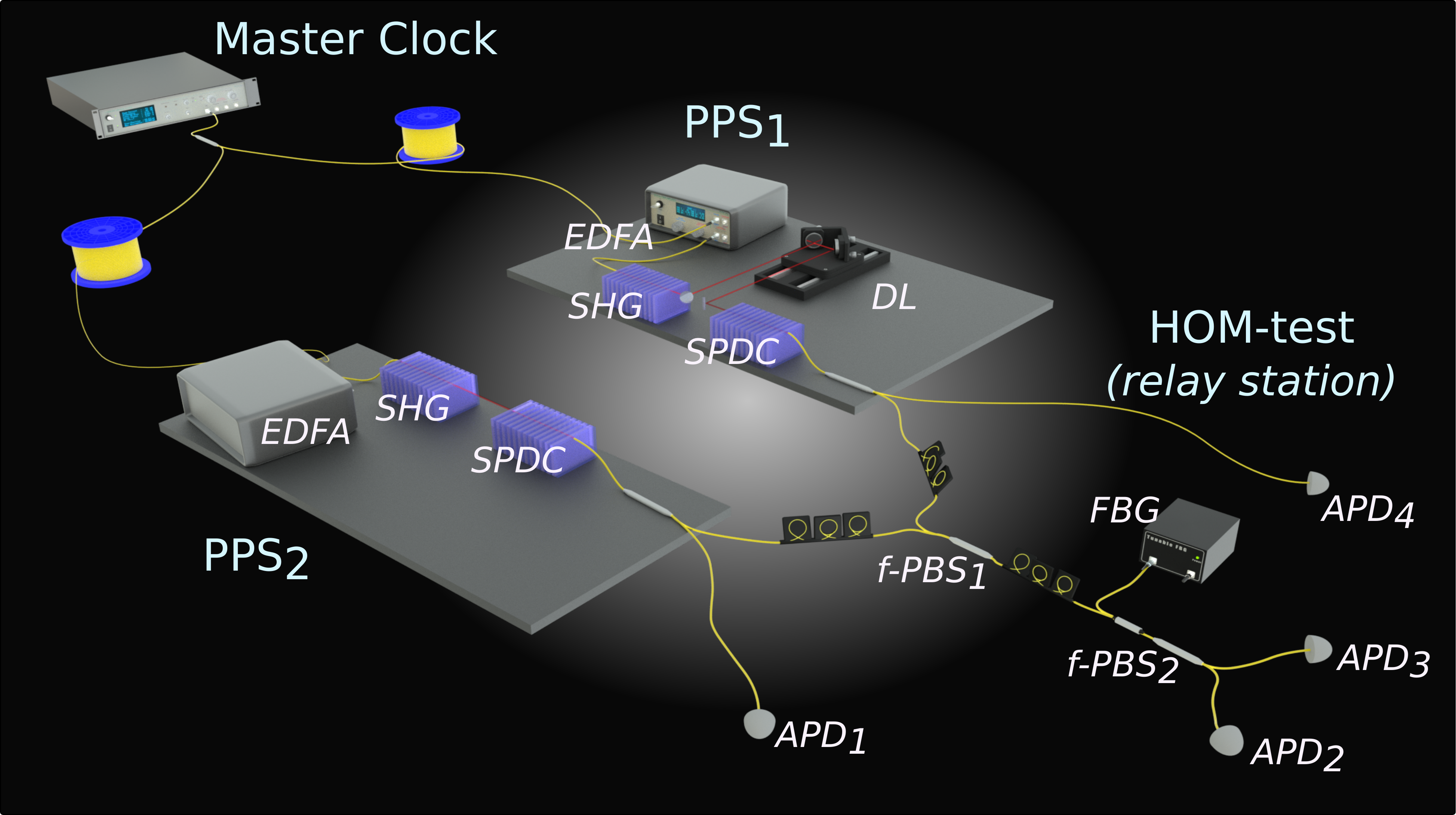}
\caption{Experimental setup employed for testing the synchronisation of two remote photon pair sources (PPSs). Conceptually, this setup is similar to those used in entanglement teleportation experiments. Optical pulses at 1540\,nm from the all-optical master clock are distributed via optical fibres to two PPSs. At the PPS nodes, pulses are locally amplified (EDFA), frequency doubled (SHG) and used to pump the photon pair generation stages (SPDC). The SPDC and the SHG stages are packaged into in-house boxes so as to reduce thermal fluctuations and their temperature is actively stabilized. At the output of the SPDC, the paired photons are separated as a function of their frequency by a fibre wavelength demultiplexer (WDM) and the source synchronisation is qualified via a HOM experiment between the photons at 1536.27\,nm heralded by their twins at 1543.73\,nm. In order to obtain the HOM dip, a delay line at PPS$_1$ allows changing, in a controlled way, the arrival time of one of the two interfering photon at the relay station. Detector APD$_2$ corresponds to 4 free-running avalanche photo-diode (IDQ 220) whose outputs are connected to a logic OR gate. They each feature a quantum detection efficiency of 25\%, 7\,$\mu$s dead time, and a dark count probability of 10$^{-6}$/ns. APD$_2$ provides a trigger signal to the three other APDs (IDQ 210) operated in a gated mode and all showing 20\% quantum detection efficiency, 7\,$\mu$s dead time, and a dark count probability of 10$^{-5}$/ns. APD's saturation rate is $\simeq$60\,kHz for all individual APDs and $\leq$200\,kHz for the multiplexed detector. These values impose a limitation to the mean number of photon pairs that are generated by each source and, in turns, to the maximum pump power at the SDPC input. Typical employed pump powers at 770\,nm are $\sim$15\,mW.}
\label{figsetup}
\end{figure*}

To demonstrate the validity of our strategy, we apply it to the synchronisation of two remote photon pair sources (PPSs) in a quantum relay configuration (see \figurename{~\ref{figsetup}}). Such an architecture lies at the heart of reliable quantum networks, because synchronous and independent entangled photon sources and relay nodes are key elements to any quantum teleportation-based optical quantum link~\cite{Sun2017,Valivarthi2016}. In our study, the all-optical synchronisation is provided by a commercial ultra-fast optical clock (\textit{Pritel}-UOC) emitting 2\,ps long pulses at 1540\,nm with a timing jitter $\leq$100\,fs. In order to obtain a quantitative measurement of the PPS synchronisation, we use the genuine quantum optical method referred to as Hong Ou Mandel (HOM) distinguishability test between independent photons coming each from one of the two sources~\cite{Abooussouan2010}. The interference visibility provides the relevant witness of the synchronisation quality. A perfect photon coalescence, i.e. a visibility of 100\%, corresponds to two interfering photons indistinguishable in the temporal modes in which they are emitted and synchronised with an accuracy better than their coherence time, namely, their intrinsic time uncertainty~\cite{Abooussouan2010}. Conversely, any timing-jitter accumulated during the photon generation process or through the quantum channel would induce a random fluctuation around the dip minimum and would result in decreased visibility. 

\begin{figure}
\includegraphics[width=0.48\textwidth]{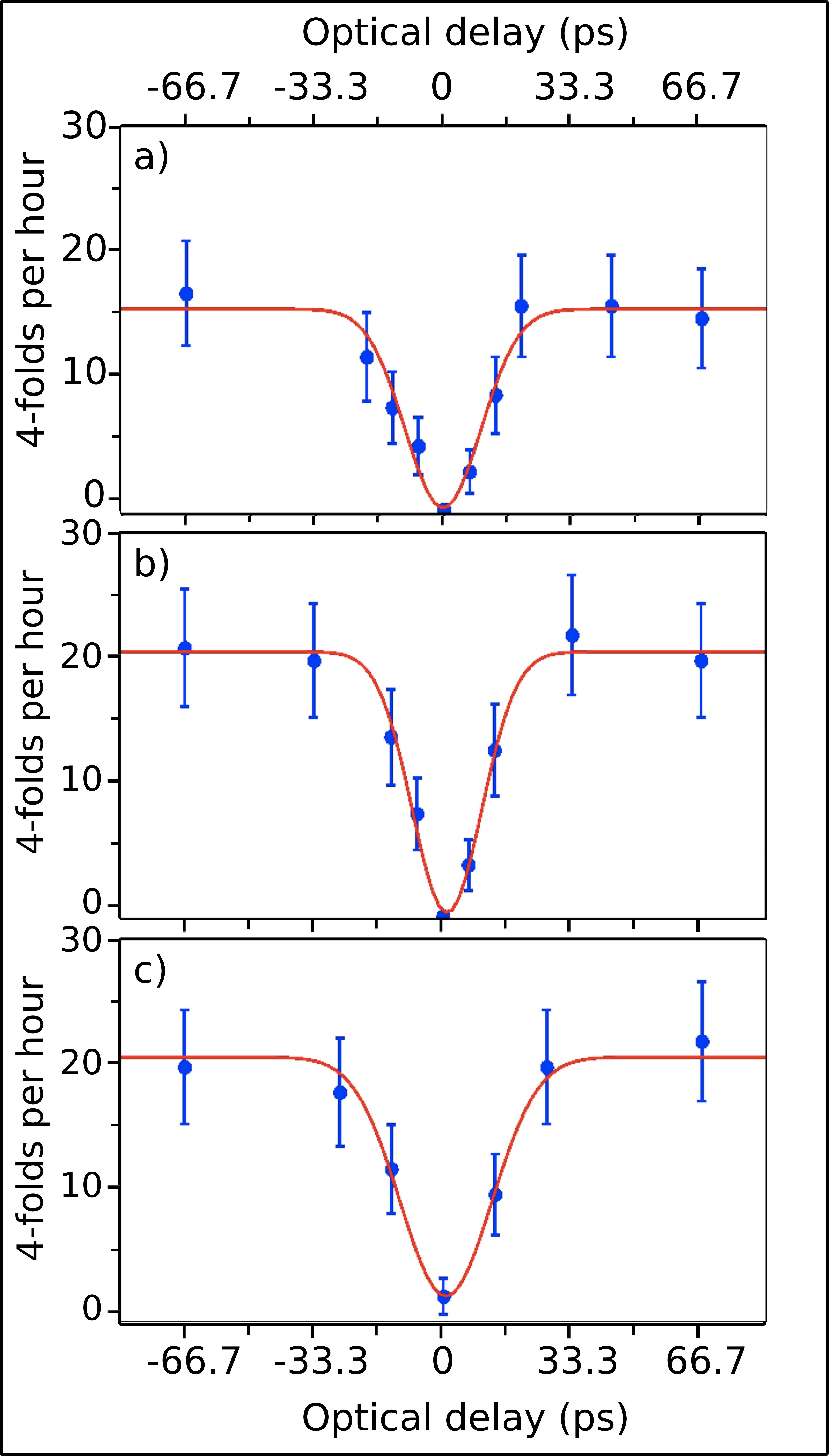}
\caption{Raw experimental results of the HOM interference between photons emitted by PPS$_1$ and PPS$_2$. The curves show 4-fold coincidences between the 2 heralding detectors and the 2 detectors at the relay station. Error bars are given as the standard deviation of Poissonian distributions. a) HOM dip for the case of two PPSs spaced by 10\,m with the master clock put perfectly halfway; b) HOM dip for an asymmetric configuration with a length mismatch $\Delta L \simeq$400\,m $\geq L_c$; c) HOM dip for the case of two PPSs spaced by 100\,km with the master clock at a distance of approximately 50$\pm$1\,km from each source. Red lines represent Gaussian fits on the experimental data. We underline higher 4-fold coincidence rates could be achieved by simply replacing the APDs by superconducting single photon detectors exhibiting high saturation rates in the MHz regime and detection efficiencies up to 90\%~\cite{Kaltenbaek2009,Benedick2012} and thus allowing to operate the experiment at higher rates.}
\label{figresults}
\end{figure}

We qualify our synchronisation scheme by considering different separations between the PPSs, including asymmetric configurations where, as in real-world configurations, the distances between the clock laser and each of the PPS are unbalanced. Due to the importance of the master clock laser characteristics for the entire setup, we start by focusing only on the relevance of its coherence properties on the synchronisation quality and temporarily put aside supplementary difficulties linked to long-distance operation. By means of preliminary classical interference (not represented), we first measured the master clock coherence length to be of $L_c\simeq$200\,m. Subsequently, we perform HOM experiments by considering the case of a symmetric configuration in which the master clock laser is set perfectly halfway between the two sources, spaced $\simeq$10\,m and that of an asymmetric one in which the distance between the clock laser and each of the EEPS shows a distance mismatch of $\Delta L\simeq$400\,m\,$>L_c$. \figurename{~\ref{figresults}}-a and -b report the HOM dips that were measured in the two operation conditions. As can be seen, there is no difference between the two experimentally obtained results and, in both configurations, the HOM profile is recovered with an extremely high raw visibility of 100\%. As accidental 4-fold coincidences between dark counts, or dark counts and single photons, are negligible, this measured raw HOM visibility coincides with its net value corrected for detection noise~\cite{Abooussouan2010}. The excellent result on visibility conceptually confirms the high quality of our synchronisation protocol and, at the same time, validates the lack of possible constraints related to the laser coherence or its distance with respect to the sources.

As a final remark, we observe that by means of error propagation, we can estimate, from our experimental data, a lower bound to HOM visibility of 95.5\%. Under the hypothesis of Gaussian distributed timing uncertainties, this value sets our minimum detectable jitters to 3\,ps. A better accuracy would be obtained with a richer statistic of 4-fold events. Without changing the pumping conditions, and, as a consequence, the photon-pair emission statistics, this can be done by using a clock with a higher repetition rate of 10\,GHz and state-of-the-art superconducting detector with ultra-low detection jitters and quantum efficiency close to 90\%~\cite{Miki2013}. This would increase the overall count by a factor 40 and lead to a measured jitter $\sim$0.5\,ps.

\subsection*{Long distance operation}

Subsequently, in order to check the validity of our method for practical network purposes, we perform the synchronisation of two PPSs spaced by 100\,km, i.e. each at 50\,km$\pm$1\,km from the master clock location. The tests are effectuated in-the-laboratory, by employing 50\,km long spools of optical fibres. Nevertheless, this configuration reproduces a realistic long-distance operation where the distribution of the maser-clock pulses over optical fibres is affected by strong propagation losses and non-negligible chromatic dispersion effects. Moreover, such a protocol must cope with thermal fluctuations in the long fibre links.

In our realization, the clock pulses directed toward each source have a mean power of $\sim$1.25\,mW. This optical power is attenuated by propagation losses in the fibres (-10\,dB) and extra-losses at the chromatic dispersion modules (-6\,dB), thus leading to a mean power of 25\,$\mu$W. At the same time, due to chromatic dispersion, the duration of the clock pulses broadens from 2\,ps to 0.85\,ns. As already stated, to face these effects, our scheme fully benefits from standard telecom technology and both attenuation and chromatic dispersion are efficiently managed thanks to plug-and-play EDFA and chromatic dispersion compensation modules based on adapted dispersion-shifted fibres~\cite{Aktas2016}. Accordingly, the most critical point lies in the fibre refractive index temperature dependence~\cite{Leviton2006} that can modify the optical distances between the two sources. For our experimental setup, thermal variations result in a drift of the HOM dip position of up to 5\,mm/h. This figure has to be compared with a dip full-width-half-maximum of $\simeq$6\,mm and with typical measurement integration time of 1\,hour/point. This effect can seriously compromise the synchronisation quality and directly maps to a shift of the HOM dip position in real time. Nevertheless, we stress that, compared to fast and random noise due to optoelectronic time-jitters, thermal fluctuations induce slow variations, an issue that has been comprehensively addressed in the past and which can be easily corrected with a standard-performance fibre-length tracking system~\cite{Minar2008}.

In order to implement an active length stabilisation system, we use the fibre spools that connect the two PPSs to the master clock as the two arms of a classical Mach-Zehnder interferometer and employ a fraction of the intensity of the master clock laser to track the position of the classical interference fringes.

The system corrects length fluctuations with an accuracy of $\sim$0.3\,mm, to be compared with a dip full-width-at-half-maximum of $\sim$6\,mm. We stress that this is not the state-of-the-art value for the stabilisation of long distance interferometers and higher accuracies are reported in the literature~\cite{Sun2017,Valivarthi2016,Minar2008}. As shown in \figurename{~\ref{figresults}}-c, the experimental results obtained with our correction system show a high raw visibility above 90.5\%, therefore validating our synchronisation approach for long distance operation. Note that this visibility value is compatible with our in-house servo-loop characteristics, and its low accuracy. Visibilities greater than 95\% would be straightforwardly achieved, with a more accurate correction system.

\section*{Conclusion}

In conclusion, we have presented and experimentally demonstrated a novel protocol that solely and simply relies on an all-optical master clock laser, the timing precision of which guarantees the synchronisation of a network blocks over long distance. Our concept exploits elementary off-the-shelf telecom technologies and optical frequency conversion stages to enable universal plug-and-play optical synchronisation of any quantum node. The synchronization approach proposed here is extremely versatile and allows in principle to add to a quantum network as many nodes (quantum memories, sources, etc.) as necessary and this whatever their nature is. In this sense, our synchronization method can therefore be seen as an enabling method for scaling up quantum networks. Compared to previously reported methods, we emphasise that our solution dramatically reduces the overhead for experimental resources. We validate our strategy for the synchronisation of distant photon pair sources based on the SPDC process by observing extremely high quality two-photon interference over 100\,km. We stress that, provided suitable nonlinear optical stages are available, our strategy is compatible with generic optical devices at the nodes of quantum networks, thus providing a relevant tool for future practical architectures.

\section*{Materials \& Methods}

\subsection*{A quantum relay configuration}

To demonstrate the validity of our strategy, we apply it to the synchronisation of two remote photon pair sources (PPSs) in a quantum relay configuration (see \figurename{~\ref{figsetup}}). The all-optical synchronisation is provided by a commercial ultra-fast optical clock (\textit{Pritel}-UOC) emitting 2\,ps long pulses at 1540\,nm, corresponding to the standard channel 47 of the international telecommunication union (ITU) wavelength grid. Typical mean powers are $\sim$2.5\,mW. The laser works at repetition rate of 2.5\,GHz with a residual timing error on the pulses' emission times $\leq$ 100\,fs. Its output is split into two at a 50:50 fibre beam splitter and distributed via standard optical fibres to the two remote PPSs. The PPSs generate paired photons at telecom wavelengths via type-0 spontaneous parametric down conversion (SPDC) in in-house periodically poled lithium niobate waveguides~\cite{Alibart2016,Abooussouan2010,Ngah2015}. Accordingly, at each PPS station, master clock pulses are amplified up to 1\,W using erbium-doped fibre amplifiers (EDFA) and frequency converted to 770\,nm via second harmonic generation (SHG) as required to use them to pump the SPDC process. The source synchronization is guaranteed only if the optical pulses pumping the remote generation processes do not exhibit relative time-jitters. This condition is automatically satisfied when the SPDC stages are directly pumped by the same master clock laser~\cite{Abooussouan2010}. Typical pump powers before the SPDC stages are set to $\sim$15\,mW at 770\,nm so as to avoid saturation at the detection stages. Correspondingly, emitted photon pairs follow a Poissonian statistics17 with mean number of pairs per pump pulse $\sim$0.0012. We note that, in view of long distance applications, time-bin entangled photons can be conveniently obtained at the output of type-0 SPDC, by locally adding on the path of pump pulses at 770\,nm an unbalanced interferometer~\cite{Xia2018,Alibart2016}. Alternatively, polarization entanglement can be obtained by replacing our non-linear waveguides with different ones allowing for a type II phase matched SPDC~\cite{Xia2018}.

Following a typical teleportation scheme, signal photons at 1543.73\,nm (ITU channel 42) coming from the PPSs are routed by a fibre wavelength de-multiplexing stage (WDM)~\cite{Alibart2016,Abooussouan2010} towards distant communication partners to be locally detected by two avalanche photodiodes (APD$_1$ and APD$_4$ in \figurename{~\ref{figsetup}}). Idler photons at 1536.27\,nm (ITU channel 50) are directed towards a central relay station where their synchronisation is evaluated (relay station in the figure). Signal and idler photons are filtered so as to have a spectral bandwidth of 800\,nm (100\,GHz) and 200\,nm (25\,GHz), corresponding to coherence times of 4\,ps and 17\,ps, respectively. Note that, in the context of quantum communication over long distances, the ps-regime has been demonstrated to enable near perfect two-photon interference while being fully compatible with standard telecom components and optical filters~\cite{Abooussouan2010}.
To evaluate the quality of the synchronization, we use HOM distinguishability test between independent photons coming each from one of the two sources~\cite{Loudon2000}. The visibility of the associated two-photon interference pattern represents a pertinent figure-of-merit assessing the quality of the synchronization. To this end, we follow the very standard approach of quantum relays~\cite{Sun2016,Valivarthi2016,Abooussouan2010} and make both our PPSs work as heralded single photon sources. In principle, this strategy allows optimizing the performances of the HOM test in terms of two-photon interference visibility. In our experiment, detection signals of photons at 1543.73\,nm at APD$_1$ and APD$_4$ announce the presence of heralded idler photons at 1536.27\,nm from PPS$_1$ and PPS$_2$, respectively~\cite{Ngah2015}.

\subsection*{A robust Hong Ou Mandel setup}

\begin{figure}
\includegraphics[width=0.5\textwidth]{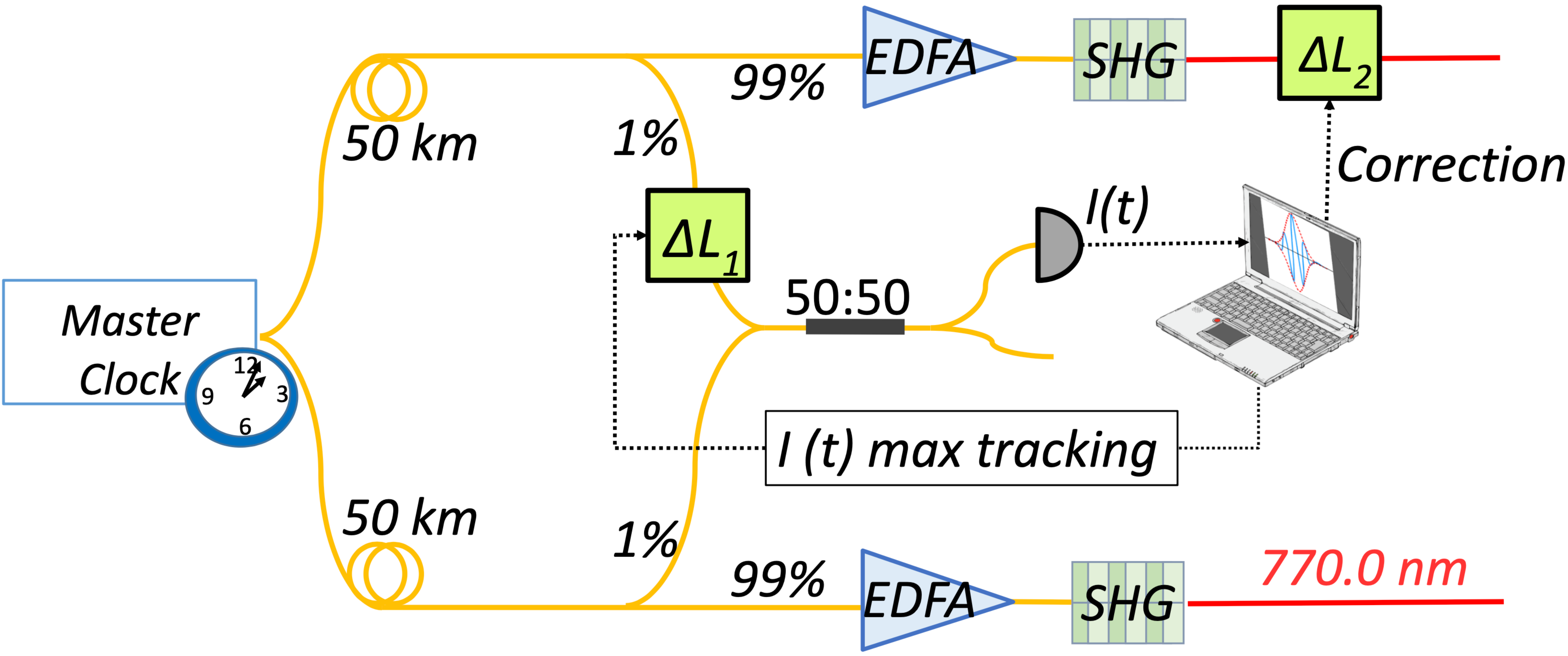}
\caption{Fibre length stabilisation setup for the HOM test in the case of two PPS spaced by 100\,km. The delay line $\Delta L_1$ allows us to check the maximum of the Mach-Zehnder classical interference fringes. In case of unwanted shift, a correction signal is sent to a second delay line, $\Delta L_2$ in order to conveniently adjust the HOM dip scan. The stabilisation system is in-house with off-the-shelf and low-cost standard electronic components.}
\label{figstab}
\end{figure}

The HOM test setup interference employed in our experiment relies on polarization-based two-photon interference~\cite{Valivarthi2016}. The heralded photons from the two sources are sent to a fibre polarisation beam splitter (f-PBS), which cleans their respective polarisations and projects their original state on the quantum state 
$|\Psi\rangle = |H\rangle |V\rangle$.
A polarisation controller (PC) is used to rotate the polarisation state of the pair by 45$^{\circ}$, leading to the state:
$|\Psi\rangle_{PC} = \frac{1}{\sqrt{2}} \left ( |H\rangle |V\rangle +|V\rangle |V\rangle - |H\rangle |H\rangle -|V\rangle |H\rangle \right )$.
The PC output is subsequently sent to a second f-PBS whose outputs are each connected to an APD (APD$_2$ and APD$_3$). The relative photon arrival times are changed by a dedicated optical delay-line (DL) and the coincidences between the counts of four APDs are recoded as a function of the relative delay. If the photons are perfectly indistinguishable in all degrees of freedom, namely spatial, spectral and temporal modes, except for their polarisation upon impinging the first f-PBS, quantum theory predicts that contributions $|H\rangle |V\rangle$ and $|V\rangle |H\rangle$ interfere destructively. As a consequence, when the photon delay is set to zero and a high-quality and stable source synchronization is achieved, the quantum state after the action of both the f-PBS leads to the state
$|\Psi\rangle_{HOM} = \frac{1}{\sqrt{2}} \left ( |V\rangle |V\rangle  - |H\rangle |H\rangle \right )$ and detectors APD$_2$ and APD$_3$ can never fire together. A dip in the four-fold coincidences among detection signals from all the APDs is thus observed. 
The interference visibility provides the relevant witness of the synchronisation quality. We recall that the success of any teleportation protocol critically depends on two-photon interference~\cite{Halder2007,Ngah2015}, thus making the exploitation of a HOM-type interferometers an extremely pertinent testbed to verify the practical capacity of any synchronisation technique. At the same time, compared with the original HOM scheme where single photons are mixed on a 50:50 beam-splitter, as discussed in Ref.~\cite{Kaiser2012}, our chosen configuration offers two main advantages : a) as the first f-PBS acts as a polarisation filter, any polarising rotation of the impinging photons only decreases the rate of detected counts but does not affect the interference visibility; b) spectral filtering of signal photons originating from separate SPDC sources is performed by a single fibre Bragg grating (FBG) filter, placed immediately after the first f-PBS. Accordingly, the indistinguishability in terms of spectral mode is automatically ensured and it is immune to the shifting wavelength of the FBG filter. Note that the bandwidth of photons at 1536.27\,nm (25\,GHz) has been chosen so as to have pure heralded single photon states, despite parasitic frequency correlations due to the SPDC process~\cite{Abooussouan2010,Ngah2015}. For the test over long distances, in order to implement an active length stabilisation system, we use the fibre spools that connect the two PPSs to the master clock as the two arms of a classical Mach-Zehnder interferometer and employ a fraction of the intensity of the master clock laser to track the position of the classical interference fringes (see \figurename{~\ref{figstab}}).
We correct optical length variations dynamically by adjusting the optical delay line used for the HOM interference in response to the variations. We note that in our experiment, the Mach-Zehnder output beam coupler combines optical signals immediately before the spontaneous parametric down conversion (SPDC) stages. In a real-world configuration the output beam splitter can be, for instance, set at the relay station, so as to take into account also length fluctuations and polarisation dispersion in the fibres downstream from the PPSs. Other configurations can rely on Michelson-like interferometers, where a fraction of the optical clock counter-propagates back and forth through each of the spools and provides information on their optical length.

\section*{Acknowledgments}

This work has been conducted within the framework of the OPTIMAL project, funded by the European Union by means of the Fond Europ\'een de d\'eveloppement regional (FEDER). The authors acknowledge financial support from Agence Nationale de la Recherche (ANR) through the Conneqt (ANR-2011-EMMA-0002), Hylight (ANR-17-CE30-0006-01) and SPOCQ (ANR-14-CE32-0019) projects, Conseil R\'egional PACA through the Distance project (Apex 2011: 2001-06938), CNRS through Conneqt (PEPS INSIS 2011), the Majlis Amanah Rakyat (MARA), Universit\'e C\^ote d'Azur (SPEED, CSI 2015) and the French government through its Investments for the Future program under the Universit\'e C\^ote d'Azur UCA-JEDI project (Quantum@UCA) managed by the ANR (ANR-15-IDEX-01). The authors acknowledge technical support from DRAKA (Prysmian group) and IDQuantique.

The illustrated synchronisation scheme has been secured by an international patent (CNRS patent N. FR11/58857 du 30/09/2011 already delivered in Europe, USA and Japan, pending in China). Authors would like to thank Anthony Martin for stimulating discussions and Fabien K\'ef\'elian for the analysis of stability of our master clock laser.

\section*{Author Contribution}

Virginia D'Auria has led the project since its conception and supervised all experiments. Bruno Fedrici has conducted most of the experimental work including the HOM measurement. Lufti Ngah and Laurent Labont\'e have worked to the nonlinear optical stage optimisation, and Florian Kaiser and Olivier Alibart on the detection and stabilisation of the experimental setup. S\'ebastien Tanzilli conceived the synchronisation scheme and co-supervised the project with Virginia D'Auria.


\begin{thebibliography}{34}
\expandafter\ifx\csname natexlab\endcsname\relax\def\natexlab#1{#1}\fi
\expandafter\ifx\csname bibnamefont\endcsname\relax
  \def\bibnamefont#1{#1}\fi
\expandafter\ifx\csname bibfnamefont\endcsname\relax
  \def\bibfnamefont#1{#1}\fi
\expandafter\ifx\csname citenamefont\endcsname\relax
  \def\citenamefont#1{#1}\fi
\expandafter\ifx\csname url\endcsname\relax
  \def\url#1{\texttt{#1}}\fi
\expandafter\ifx\csname urlprefix\endcsname\relax\def\urlprefix{URL }\fi
\providecommand{\bibinfo}[2]{#2}
\providecommand{\eprint}[2][]{\url{#2}}

\bibitem[{\citenamefont{Kimble}(2008)}]{Kimble2008}
\bibinfo{author}{\bibfnamefont{H.~J.} \bibnamefont{Kimble}},
  \bibinfo{journal}{Nature} \textbf{\bibinfo{volume}{\textbf{453}}},
  \bibinfo{pages}{1023} (\bibinfo{year}{2008}).

\bibitem[{\citenamefont{Simon}(2017)}]{Simon2017}
\bibinfo{author}{\bibfnamefont{C.}~\bibnamefont{Simon}}, \bibinfo{journal}{Nat.
  Photonics} \textbf{\bibinfo{volume}{\textbf{11}}}, \bibinfo{pages}{678}
  (\bibinfo{year}{2017}).

\bibitem[{\citenamefont{Wehner et~al.}(2018)\citenamefont{Wehner, Elkouss, and
  Hanson}}]{Wehner2018}
\bibinfo{author}{\bibfnamefont{S.}~\bibnamefont{Wehner}},
  \bibinfo{author}{\bibfnamefont{D.}~\bibnamefont{Elkouss}}, \bibnamefont{and}
  \bibinfo{author}{\bibfnamefont{R.}~\bibnamefont{Hanson}},
  \bibinfo{journal}{Science} \textbf{\bibinfo{volume}{\textbf{362}}},
  \bibinfo{pages}{eaam9288} (\bibinfo{year}{2018}).

\bibitem[{\citenamefont{Saglamyurek~\textit{et al.}}(2015)}]{Saglamyurek2015}
\bibinfo{author}{\bibfnamefont{E.}~\bibnamefont{Saglamyurek~\textit{et al.}}},
  \bibinfo{journal}{Nat. Photonics} \textbf{\bibinfo{volume}{\textbf{9}}},
  \bibinfo{pages}{83} (\bibinfo{year}{2015}).

\bibitem[{\citenamefont{Parigi~\textit{et al.}}(2015)}]{Parigi2015}
\bibinfo{author}{\bibfnamefont{V.}~\bibnamefont{Parigi~\textit{et al.}}},
  \bibinfo{journal}{Nat. Commun.} \textbf{\bibinfo{volume}{\textbf{6}}},
  \bibinfo{pages}{7706} (\bibinfo{year}{2015}).

\bibitem[{\citenamefont{Woerkom~\textit{et al.}}(2018)}]{Woerkom2018}
\bibinfo{author}{\bibfnamefont{D.~J.} \bibnamefont{Woerkom~\textit{et al.}}},
  \bibinfo{journal}{Phys. Rev. X} \textbf{\bibinfo{volume}{\textbf{8}}},
  \bibinfo{pages}{041018} (\bibinfo{year}{2018}).

\bibitem[{\citenamefont{Maring~\textit{et al.}}(2017)}]{Maring2017}
\bibinfo{author}{\bibfnamefont{N.}~\bibnamefont{Maring~\textit{et al.}}},
  \bibinfo{journal}{Nature} \textbf{\bibinfo{volume}{\textbf{551}}},
  \bibinfo{pages}{485} (\bibinfo{year}{2017}).

\bibitem[{\citenamefont{Kaiser~\textit{et al.}}(2015)}]{Kaiser2015}
\bibinfo{author}{\bibfnamefont{F.}~\bibnamefont{Kaiser~\textit{et al.}}},
  \bibinfo{journal}{IEEE J. Sel. Top. Quantum Electron.}
  \textbf{\bibinfo{volume}{\textbf{21}}}, \bibinfo{pages}{69}
  (\bibinfo{year}{2015}).

\bibitem[{\citenamefont{Kaiser~\textit{et al.}}(2019)}]{Kaiser2019}
\bibinfo{author}{\bibfnamefont{F.}~\bibnamefont{Kaiser~\textit{et al.}}},
  \bibinfo{journal}{Opt. Express} \textbf{\bibinfo{volume}{\textbf{27}}},
  \bibinfo{pages}{25603} (\bibinfo{year}{2019}).

\bibitem[{\citenamefont{Sun~\textit{et al.}}(2017)}]{Sun2017}
\bibinfo{author}{\bibfnamefont{Q.-C.} \bibnamefont{Sun~\textit{et al.}}},
  \bibinfo{journal}{Optica} \textbf{\bibinfo{volume}{\textbf{4}}},
  \bibinfo{pages}{1214} (\bibinfo{year}{2017}).

\bibitem[{\citenamefont{Sun~\textit{et al.}}(2016)}]{Sun2016}
\bibinfo{author}{\bibfnamefont{Q.-C.} \bibnamefont{Sun~\textit{et al.}}},
  \bibinfo{journal}{Nat. Photonics} \textbf{\bibinfo{volume}{\textbf{10}}},
  \bibinfo{pages}{671} (\bibinfo{year}{2016}).

\bibitem[{\citenamefont{Valivarthi~\textit{et al.}}(2016)}]{Valivarthi2016}
\bibinfo{author}{\bibfnamefont{R.}~\bibnamefont{Valivarthi~\textit{et al.}}},
  \bibinfo{journal}{Nat. Photonics} \textbf{\bibinfo{volume}{\textbf{10}}},
  \bibinfo{pages}{676} (\bibinfo{year}{2016}).

\bibitem[{\citenamefont{Jin et~al.}(2015)\citenamefont{Jin, Takeoka, Takagi,
  Shimizu, and Sasaki}}]{Jin2015}
\bibinfo{author}{\bibfnamefont{R.-B.} \bibnamefont{Jin}},
  \bibinfo{author}{\bibfnamefont{M.}~\bibnamefont{Takeoka}},
  \bibinfo{author}{\bibfnamefont{U.}~\bibnamefont{Takagi}},
  \bibinfo{author}{\bibfnamefont{R.}~\bibnamefont{Shimizu}}, \bibnamefont{and}
  \bibinfo{author}{\bibfnamefont{M.}~\bibnamefont{Sasaki}},
  \bibinfo{journal}{Sci. Rep.} \textbf{\bibinfo{volume}{\textbf{5}}},
  \bibinfo{pages}{9333} (\bibinfo{year}{2015}).

\bibitem[{\citenamefont{Xia et~al.}(2018)\citenamefont{Xia, Sun, Zhang, and
  Pan}}]{Xia2018}
\bibinfo{author}{\bibfnamefont{X.-X.} \bibnamefont{Xia}},
  \bibinfo{author}{\bibfnamefont{Q.-C.} \bibnamefont{Sun}},
  \bibinfo{author}{\bibfnamefont{Q.}~\bibnamefont{Zhang}}, \bibnamefont{and}
  \bibinfo{author}{\bibfnamefont{J.-W.} \bibnamefont{Pan}},
  \bibinfo{journal}{Quantum Sci. Technol.}
  \textbf{\bibinfo{volume}{\textbf{3}}}, \bibinfo{pages}{014012}
  (\bibinfo{year}{2018}).

\bibitem[{\citenamefont{Tao~\textit{et al.}}(2006)}]{Tao2006}
\bibinfo{author}{\bibfnamefont{Y.}~\bibnamefont{Tao~\textit{et al.}}},
  \bibinfo{journal}{Phys. Rev. Lett.} \textbf{\bibinfo{volume}{\textbf{96}}},
  \bibinfo{pages}{110501} (\bibinfo{year}{2006}).

\bibitem[{\citenamefont{Halder~\textit{et al.}}(2007)}]{Halder2007}
\bibinfo{author}{\bibfnamefont{M.}~\bibnamefont{Halder~\textit{et al.}}},
  \bibinfo{journal}{Nat. Phys.} \textbf{\bibinfo{volume}{\textbf{3}}},
  \bibinfo{pages}{692} (\bibinfo{year}{2007}).

\bibitem[{\citenamefont{Alibart~\textit{et al.}}(2016)}]{Alibart2016}
\bibinfo{author}{\bibfnamefont{O.}~\bibnamefont{Alibart~\textit{et al.}}},
  \bibinfo{journal}{J. Opt.} \textbf{\bibinfo{volume}{\textbf{18}}},
  \bibinfo{pages}{104001} (\bibinfo{year}{2016}).

\bibitem[{\citenamefont{Kaltenbaek~\textit{et al.}}(2009)}]{Kaltenbaek2009}
\bibinfo{author}{\bibfnamefont{R.}~\bibnamefont{Kaltenbaek~\textit{et al.}}},
  \bibinfo{journal}{Phys. Rev. A} \textbf{\bibinfo{volume}{\textbf{79}}},
  \bibinfo{pages}{040302} (\bibinfo{year}{2009}).

\bibitem[{\citenamefont{Shelton~\textit{et al.}}(2001)}]{Shelton2001}
\bibinfo{author}{\bibfnamefont{R.~K.} \bibnamefont{Shelton~\textit{et al.}}},
  \bibinfo{journal}{Science} \textbf{\bibinfo{volume}{\textbf{293}}},
  \bibinfo{pages}{1286} (\bibinfo{year}{2001}).

\bibitem[{\citenamefont{Cundiff et~al.}(2003)\citenamefont{Cundiff, Kolner,
  Corkum, Diddams, and Telle}}]{Cundiff2003}
\bibinfo{author}{\bibfnamefont{S.}~\bibnamefont{Cundiff}},
  \bibinfo{author}{\bibfnamefont{B.}~\bibnamefont{Kolner}},
  \bibinfo{author}{\bibfnamefont{P.}~\bibnamefont{Corkum}},
  \bibinfo{author}{\bibfnamefont{S.}~\bibnamefont{Diddams}}, \bibnamefont{and}
  \bibinfo{author}{\bibfnamefont{H.}~\bibnamefont{Telle}},
  \bibinfo{journal}{IEEE J. Sel. Top. Quantum Electron.}
  \textbf{\bibinfo{volume}{\textbf{9}}}, \bibinfo{pages}{969}
  (\bibinfo{year}{2003}).

\bibitem[{\citenamefont{Halder~\textit{et al.}}(2008)}]{Halder2008}
\bibinfo{author}{\bibfnamefont{M.}~\bibnamefont{Halder~\textit{et al.}}},
  \bibinfo{journal}{New J. Phys.} \textbf{\bibinfo{volume}{\textbf{10}}},
  \bibinfo{pages}{023027} (\bibinfo{year}{2008}).

\bibitem[{\citenamefont{Hadfield}(2009)}]{Hadfield2009}
\bibinfo{author}{\bibfnamefont{R.~H.} \bibnamefont{Hadfield}},
  \bibinfo{journal}{New J. Phys.} \textbf{\bibinfo{volume}{\textbf{3}}},
  \bibinfo{pages}{696} (\bibinfo{year}{2009}).

\bibitem[{\citenamefont{Aboussouan et~al.}(2010)\citenamefont{Aboussouan,
  Alibart, Ostrowsky, Baldi, and Tanzilli}}]{Abooussouan2010}
\bibinfo{author}{\bibfnamefont{P.}~\bibnamefont{Aboussouan}},
  \bibinfo{author}{\bibfnamefont{O.}~\bibnamefont{Alibart}},
  \bibinfo{author}{\bibfnamefont{D.~B.} \bibnamefont{Ostrowsky}},
  \bibinfo{author}{\bibfnamefont{P.}~\bibnamefont{Baldi}}, \bibnamefont{and}
  \bibinfo{author}{\bibfnamefont{S.}~\bibnamefont{Tanzilli}},
  \bibinfo{journal}{Phys. Rev. A} \textbf{\bibinfo{volume}{\textbf{81}}},
  \bibinfo{pages}{021801} (\bibinfo{year}{2010}).

\bibitem[{\citenamefont{Ngah et~al.}(2015)\citenamefont{Ngah, Alibart,
  Labont\'e, D'Auria, and Tanzilli}}]{Ngah2015}
\bibinfo{author}{\bibfnamefont{L.~A.} \bibnamefont{Ngah}},
  \bibinfo{author}{\bibfnamefont{O.}~\bibnamefont{Alibart}},
  \bibinfo{author}{\bibfnamefont{L.}~\bibnamefont{Labont\'e}},
  \bibinfo{author}{\bibfnamefont{V.}~\bibnamefont{D'Auria}}, \bibnamefont{and}
  \bibinfo{author}{\bibfnamefont{S.}~\bibnamefont{Tanzilli}},
  \bibinfo{journal}{Laser Photon. Rev.} \textbf{\bibinfo{volume}{\textbf{9}}},
  \bibinfo{pages}{L1} (\bibinfo{year}{2015}).

\bibitem[{\citenamefont{Aktas~\textit{et al.}}(2016)}]{Aktas2016}
\bibinfo{author}{\bibfnamefont{D.}~\bibnamefont{Aktas~\textit{et al.}}},
  \bibinfo{journal}{Laser Photon. Rev.} \textbf{\bibinfo{volume}{\textbf{10}}},
  \bibinfo{pages}{451} (\bibinfo{year}{2016}).

\bibitem[{\citenamefont{Agrawal}(2006)}]{Agrawal2006}
\bibinfo{author}{\bibfnamefont{G.~P.} \bibnamefont{Agrawal}},
  \bibinfo{journal}{Academic Press, Boston}  (\bibinfo{year}{2006}).

\bibitem[{\citenamefont{Benedick et~al.}(2012)\citenamefont{Benedick, Fujimoto,
  and K\"artner}}]{Benedick2012}
\bibinfo{author}{\bibfnamefont{A.~J.} \bibnamefont{Benedick}},
  \bibinfo{author}{\bibfnamefont{J.~G.} \bibnamefont{Fujimoto}},
  \bibnamefont{and} \bibinfo{author}{\bibfnamefont{F.~X.}
  \bibnamefont{K\"artner}}, \bibinfo{journal}{Nat. Photonics}
  \textbf{\bibinfo{volume}{\textbf{6}}}, \bibinfo{pages}{97}
  (\bibinfo{year}{2012}).

\bibitem[{\citenamefont{Rozpedek~\textit{et al.}}(2019)}]{Rozpedek2019}
\bibinfo{author}{\bibfnamefont{F.}~\bibnamefont{Rozpedek~\textit{et al.}}},
  \bibinfo{journal}{Phys. Rev. A} \textbf{\bibinfo{volume}{\textbf{99}}},
  \bibinfo{pages}{052330} (\bibinfo{year}{2019}).

\bibitem[{\citenamefont{Lisdat~\textit{et al.}}(2016)}]{Lisdat2016}
\bibinfo{author}{\bibfnamefont{C.}~\bibnamefont{Lisdat~\textit{et al.}}},
  \bibinfo{journal}{Nat. Commun.} \textbf{\bibinfo{volume}{\textbf{7}}},
  \bibinfo{pages}{12443 EP} (\bibinfo{year}{2016}).

\bibitem[{\citenamefont{Miki et~al.}(2013)\citenamefont{Miki, Yamashita, Terai,
  and Wang}}]{Miki2013}
\bibinfo{author}{\bibfnamefont{S.}~\bibnamefont{Miki}},
  \bibinfo{author}{\bibfnamefont{T.}~\bibnamefont{Yamashita}},
  \bibinfo{author}{\bibfnamefont{H.}~\bibnamefont{Terai}}, \bibnamefont{and}
  \bibinfo{author}{\bibfnamefont{Z.}~\bibnamefont{Wang}},
  \bibinfo{journal}{Opt. Express} \textbf{\bibinfo{volume}{\textbf{7}}},
  \bibinfo{pages}{21} (\bibinfo{year}{2013}).

\bibitem[{\citenamefont{Leviton and Frey}(2006)}]{Leviton2006}
\bibinfo{author}{\bibfnamefont{D.~B.} \bibnamefont{Leviton}} \bibnamefont{and}
  \bibinfo{author}{\bibfnamefont{B.~J.} \bibnamefont{Frey}},
  \bibinfo{journal}{Proc. SPIE, Optomechanical Technologies for Astronomy}
  \textbf{\bibinfo{volume}{\textbf{6273}}}, \bibinfo{pages}{62732K}
  (\bibinfo{year}{2006}).

\bibitem[{\citenamefont{Min\'a\v{r} et~al.}(2008)\citenamefont{Min\'a\v{r},
  De~Riedmatten, Simon, Zbinden, and Gisin}}]{Minar2008}
\bibinfo{author}{\bibfnamefont{J.}~\bibnamefont{Min\'a\v{r}}},
  \bibinfo{author}{\bibfnamefont{H.}~\bibnamefont{De~Riedmatten}},
  \bibinfo{author}{\bibfnamefont{C.}~\bibnamefont{Simon}},
  \bibinfo{author}{\bibfnamefont{H.}~\bibnamefont{Zbinden}}, \bibnamefont{and}
  \bibinfo{author}{\bibfnamefont{N.}~\bibnamefont{Gisin}},
  \bibinfo{journal}{Phys. Rev. A} \textbf{\bibinfo{volume}{\textbf{77}}},
  \bibinfo{pages}{052325} (\bibinfo{year}{2008}).

\bibitem[{\citenamefont{Loudon}(2000)}]{Loudon2000}
\bibinfo{author}{\bibfnamefont{R.}~\bibnamefont{Loudon}},
  \bibinfo{journal}{Oxford University Press}  (\bibinfo{year}{2000}).

\bibitem[{\citenamefont{Kaiser~\textit{et al.}}(2012)}]{Kaiser2012}
\bibinfo{author}{\bibfnamefont{F.}~\bibnamefont{Kaiser~\textit{et al.}}},
  \bibinfo{journal}{New J. Phys.} \textbf{\bibinfo{volume}{\textbf{14}}},
  \bibinfo{pages}{085015} (\bibinfo{year}{2012}).

\end{thebibliography}

\end{document}